# Hyperbolic spoof plasmons in layered equivalent graphene metasurfaces

*Li-Zheng Yin, Tie-Jun Huang, Di Wang, and Pu-Kun Liu* *

Dr. L.-Z. Yin, Dr. T.-J. Huang, Dr. D. Wang, Prof. P.-K. Liu
State Key Laboratory of Advanced Optical Communication Systems and Networks,
Department of Electronics, Peking University, Beijing 100871, China
E-mail: pkliu@pku.edu.cn

Advances in graphene plasmonics offer numerous opportunities for enabling the design and manufacture of a variety of nanoelectronics and other exciting optical devices. However, due to the limitation of material properties, its operating frequency cannot drop to the microwave range. In this work, a new concept of microwave equivalent graphene based on the ultrathin monolayer plasmonic metasurface is proposed and demonstrated. Based on this concept, elliptical and hyperbolic dispersion can be theoretically obtained by stacking the equivalent graphene metasurfaces periodically. As proofs of the concept and method, an elliptical and an all-metal hyperbolic metamaterial are designed and numerically demonstrated. As a specified realization of the method, a practical hyperbolic metamaterial is fabricated and experimentally investigated with its validity verified by the directional propagation and photonic spin Hall effect. Furthermore, to investigate the validity of the method under extreme parameter conditions, a proof-of-concept hyperlens is designed and fabricated, with its near-field resolution of 0.05λ experimentally verified. Based on the proposed concept, diverse optical graphene metamaterials such as focusing lens, dispersion-dependent directional couplers, and epsilon-near-zero materials can also be realized in the microwave regime.

## 1. Introduction

Graphene, a single two-dimension(2D) layer of carbon atoms arranged in the form of honeycomb lattice, has shown many remarkable optical, electric, mechanical, and thermal properties.[1] Among the diverse aspects, it's the optical properties of electric tunability and intrinsic plasmons [1-3] that make the graphene ubiquitous in different areas ranging from broadband perfect absorbers,[4,5] biosensors,[6,7] and ultrafast photodetector,[8,9] and so on. Compared with the conventional plasmonic materials, graphene has the advantages of high surface plasmon polaritons (SPPs) confinement and low ohmic loss.[1,3] Furthermore, the conductivity of the graphene, which is a perfect parameter to describe its optical property, can be tuned by applying external voltage because of its impact on the strong interband transitions. Based on these merits, graphene becomes a good candidate to construct hyperbolic metamaterials (HMMs) that support the electromagnetic modes whose equal frequency contours(EFCs) are the hyperbola.[10-11] The HMMs, of which the longitudinal and transverse components of the effective permittivity tensor have opposite signs, have shown enormous applications in subwavelength imaging,[12-14] focusing,[15-17] negative refraction,[18-21] and directional single-photon emission,[22,23] and so on. Compared with the mainstream approach to construct HMMs by stacking subwavelength plasmonic metal and dielectric multilayers periodically, graphene HMMs have the advantages of low ohmic loss and wide electric tunability. Recent advances in nanofabrication technology have also greatly promoted the prosperity of graphene HMMs composed of periodic subwavelength dielectric-graphene.[24-28] However, similar to the conventional plasmonic materials whose operating frequency is confined in the infrared and optical regimes only, graphene as well as the corresponding HMMs also cannot function in the microwave regime due to the limitation of the material properties.

To develop plasmonic materials at low frequency, spoof surface plasmon(SSP) which is a novel propagating surface wave supported by corrugated metal emerges.[29] After that, more and more novel structures were proposed and demonstrated to support the propagation of SSP. Among these diverse SSP waveguides, plasmonic metasurfaces stand out for their subwavelength thickness and low metal loss.[30-35] The conventional plasmonic metasurfaces are tri-layer structures consisting of flat metal mirrors and patches, with the dielectric embedded between them. For convenience, most researchers model the whole structures as volume plasmonic materials with negative permittivity or permeability according to the effective medium theory.[36] However, considering the nonignorable thickness and the only half-space SSP field distribution,[30-35] the conventional plasmonic metasurfaces with tri-layer structures cannot become the equivalent substitution of graphene in the microwave regime.

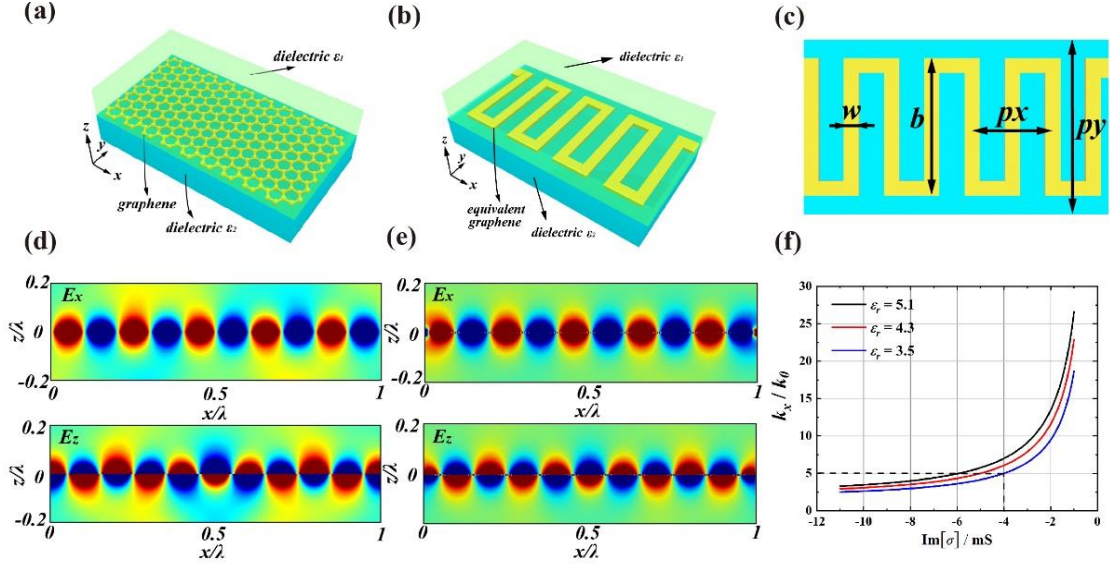

**Figure 1**. (a) and (b) are schematics of the real graphene and the proposed equivalent graphene, respectively. (c) The shape and structure parameters of the equivalent graphene. (d) and (e) are respective electric field distribution of the graphene SPP and equivalent graphene SSP. (f) The relation between propagation constant and effective conductivity of the equivalent graphene SSP with $\varepsilon_r = 3.5 - 0.01i$, $4.3 - 0.01i$, and $5.1 - 0.01i$, where the drop lines mark the operating point of electric field distribution in (d) and (e).

In this work, we propose the concept of equivalent graphene, a monolayer ultrathin plasmonic metasurface who supports the similar evanescent wave field distribution to the graphene, and analytically demonstrate that it can also be modeled as a sheet with effective conductivity $\sigma_m$. Based on this concept, we prove that elliptical and hyperbolic EFCs can be realized by stacking the equivalent graphene plasmonic metasurfaces(EGPMs) periodically. As proofs of the concept and method, we propose a snakelike metal EGPM and design an elliptical and a hyperbolic metamaterial with it, where the designed HMM is the all-metal structure qualified to work in the high power situations. This is because the negative transverse components of the effective permittivity tensor mainly depend on the structure parameters of the EGPMs according to the proposed method. Therefore, arbitrary dielectrics even air can be selected to construct the anisotropic metamaterials. The practical EFCs of the designed metamaterials obtained by the two-dimension (2D) spatial Fourier transform of the field distribution agrees well with the theory, confirming the effectiveness of the proposed method. It' noted that other monolayer plasmonic metasurfaces with diversified shapes can also be equivalent to graphene according to this concept. Compared with the anisotropic metamaterials constructed by periodic metal-dielectric, the proposed structures have the advantage of the low ohmic loss and high flexibility in design. As a specified realization, a practical hyperbolic metamaterial is fabricated and experimentally investigated with its validity verified by the directional propagation. In addition, photonic spin Hall effect (PSHE) is also experimentally observed by the polarization-controlled routing of subwavelength modes. Furthermore, to investigate the validity of the method under extreme parameter conditions, a proof-of-concept hyperlens is designed and fabricated with its near-field resolution of 0.05λ experimentally verified. Our study may open up new routes in designing plasmonics metamaterials such as the focusing lens, dispersion-dependent directional couplers, epsilon-near-zero materials, and other practical microwave devices.

**2. Concept of Equivalent Graphene Plasmonic Metasurfaces**

Monolayer graphene can be modeled as an ultrathin isotropic surface with effective conductivity $\sigma_g$ which can be calculated by Kubo formula, as illustrated by **Figure 1(a).**[37] For the sake of clarity, we only present the semiclassical Drude-like model of the effective conductivity of graphene

$$\sigma_g \cong \frac{ie^2\mu_c}{\pi\hbar^2(\omega+i\tau^{-1})}, \qquad (1)$$

where $\mu_c$ and $\tau$ represent chemical potential and electron relaxation time, respectively. According to classical electromagnetic theory, the imaginary part of the conductivity Im[$\sigma_g$] determines the transmission characteristics of the electromagnetic waves(EMWs), while the

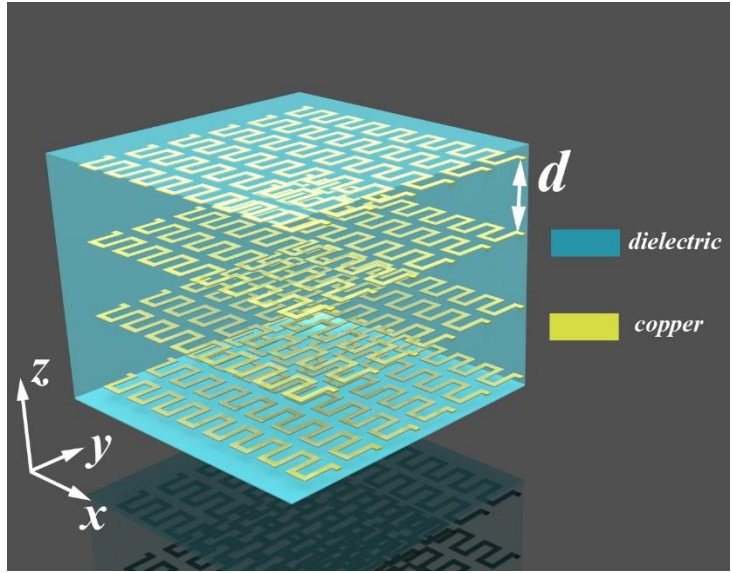

**Figure 2**. Schematic of the constructed anisotropic metamaterial. The distance between the adjacent EGPMs sheets is *d*.

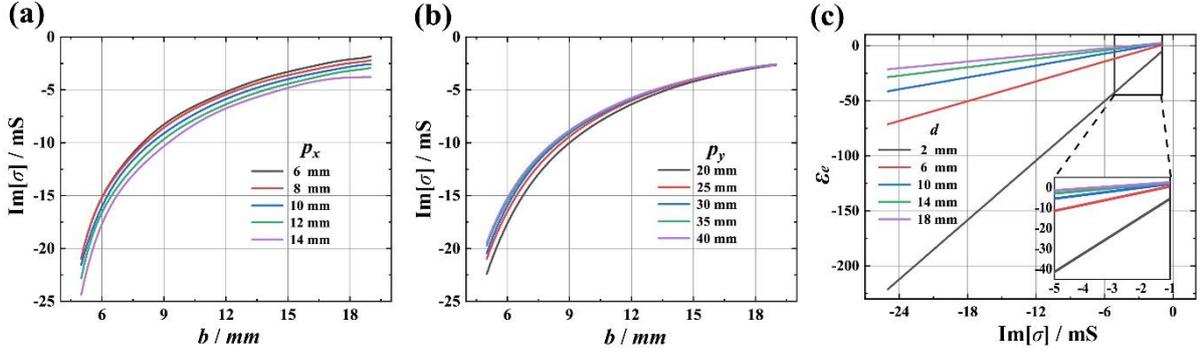

**Figure 3**. (a) and (b) show the effective conductivity of the EGPMs with different geometrical size. (a) is calculated under *w* = 2 mm, $\varepsilon_r$ = 3.5 − 0.01*i*, and $p_y$ = 30 mm, and (b) is calculated under *w* = 2 mm, $\varepsilon_r$ = 3.5 − 0.01*i*, and $p_x$ = 10 mm. (c) Effective permittivity of the constructed anisotropic metamaterials with different distance *d* and effective conductivity $\sigma_m$.

real part Re[$\sigma_g$] determines the loss.[38] Compared with Re[$\sigma_g$], the much larger Im[$\sigma_g$] in the optical frequency means that the graphene can effectively support the propagation of SPP with low loss. **Figure 1(d)** displays the typical electric field distribution of graphene with $\sigma_g$ = 0.05 − 4*i* mS and $\varepsilon_1$ = $\varepsilon_2$ =3.5 − 0.01*i*, where $\varepsilon_1$ and $\varepsilon_2$ represent the permittivity of the dielectric above and below the graphene. However, from Equation (1) we find that when the operating frequency drops to the microwave range where $\omega$ ~ $10^{10}$ rad/s, Re[$\sigma_g$] will be much larger than Im[$\sigma_g$] for graphene, which means that SPP cannot be efficiently excited. In order to fill the blank of graphene in the microwave band, we propose the concept of equivalent graphene, a monolayer ultrathin plasmonic metasurface that can support the Eigen SSP. Next, we derive the dispersion curve of the EGPMs and obtain the relationship between the effective conductivity $\sigma_m$ and propagation constant $k_x$. Here, we consider the configuration with the EGPM embedded in two different dielectrics whose relative dielectric constant is $\varepsilon_1$ and $\varepsilon_2$, respectively, as illustrated by **Figure 1(b)**. In this work, we propose a snakelike metal EGPM whose structure parameters are illustrated in **Figure 1(c)**. It's noted that other monolayer ultrathin plasmonic metasurfaces with diverse shapes can also be equivalent to graphene according to this concept. In this work, we only consider the TM mode so the electromagnetic waves above and below the ultrathin sheet have the form

$$H_{y1} = A_1 e^{-k_1 z} e^{-jk_x x}, \tag{2a}$$

$$E_{x1} = -jA_1 \frac{k_1}{\omega \varepsilon_0 \varepsilon_1} e^{-k_1 z} e^{-jk_x x}, \tag{2b}$$

$$E_{z1} = -A_1 \frac{k_x}{\omega \varepsilon_0 \varepsilon_1} e^{-k_1 z} e^{-jk_x x}, \quad \text{for } z > 0 \tag{2c}$$

$$H_{y2} = A_2 e^{k_2 z} e^{-jk_x x}, \tag{3a}$$

$$E_{x2} = jA_2 \frac{k_2}{\omega \varepsilon_0 \varepsilon_2} e^{k_2 z} e^{-jk_x x}, \tag{3b}$$

$$E_{z2} = -A_2 \frac{k_x}{\omega \varepsilon_0 \varepsilon_2} e^{k_2 z} e^{-jk_x x}. \quad \text{for } z < 0 \tag{3c}$$

where $k_i = \sqrt{k_x^2 - \varepsilon_i k_0^2}$ $i = 1,2$. $k_x$ and $k_0$ represent the wave vector in $x$ direction (propagation constant) and free space, respectively. When we model the EGPM as an ultrathin sheet with effective conductivity $\sigma_m$, the boundary conditions can be expressed as

$$E_{1x} = E_{2x}|_{z=0}, \tag{4a}$$

$$H_{2y} - H_{1y} = \sigma_m E_x|_{z=0}. \tag{4b}$$

We assume that the structure is ideally periodic in *y* direction. Substituting Equation (2) into Equation (3), we obtain the dispersion relation for TM modes

$$\frac{\varepsilon_1}{k_1} + \frac{\varepsilon_2}{k_2} = j\frac{\sigma_m}{w\varepsilon_0}. \tag{5}$$

In this work, we consider the case $\varepsilon_1 = \varepsilon_2 = \varepsilon_r$ so the dispersion relation can be simplified as

$$\sigma_m = \frac{2\omega\varepsilon_0\varepsilon_r}{j\sqrt{k_x^2 - \varepsilon_r k_0^2}}. \tag{6}$$

We plot the dispersion curves with $\varepsilon_r = 3.5 - 0.01i$, $4.3 - 0.01i$, and $5.1 - 0.01i$ in **Figure 1(f)**. To verify the effect of the equivalence, the calculated electric field distribution of the real graphene and EGPM with $\sigma_g = \sigma_m = 0.05 - 4i$ mS and $\varepsilon_r = 3.5 - 0.01i$ are illustrated in Figure 1(d) and **1(e)**, respectively. The visually consistent electric field distributions are in agreement with theory, confirming the ideal effect of the concept of equivalent graphene. The drop lines in Figure 1(f) mark the corresponding operating point. The exact structure parameters of the equivalent graphene with $\sigma_m = 0.05 - 4i$ mS at 2 GHz are $p_x = 2.5$ mm, $p_y = 10$ mm, $w = 0.63$ mm, and $b = 8.5$ mm.

**3. Anisotropic Dispersion in multilayer Equivalent Graphene Plasmonic Metasurfaces**

To investigate the relation between the dispersion characteristic and the structure parameters, we numerically calculate the effective conductivity of EGPMs with different sizes by the Eigen frequency Solver of commercial FEM simulation software COMSOL Multiphysics 5.4. The propagation constant, which is related to the effective conductivity, can be obtained by solving the Eigen frequency of the EGPMs with periodic boundary conditions. **Figure 3(a)** illustrates the effective conductivity $\sigma_m$ versus $b$, $p_x$ under $w = 2$ mm, $\varepsilon_r = 3.5 - 0.01i$, $p_y = 30$ mm, and **Figure 3(b)** illustrates the effective conductivity $\sigma_m$ versus $b$, $p_y$ under $w = 2$ mm, $\varepsilon_r = 3.5 - 0.01i$, $p_x = 10$ mm. The central operating frequency of the EGPMs is 2 GHz. From the results shown in the two figures, we can conclude that the effective conductivity $\sigma_m$ is predominantly determined by the width $b$ of the EGPMs and insensitive to the period $p_x$ and $p_y$, which offers us the possibilities to design the EGPMs with a wide range of $\sigma_m$. However, considering the Brillouin zone folding effect, it's the $p_x$ who determines the upper limit of the propagation constant.[38] When we tune the width $b$ from 5 mm to 19 mm, a relatively large range of effective conductivity from $-24.4i$ mS to $-1.8i$ mS can be obtained. This range can also be enlarged by further increasing or decreasing the width $b$. The real part of the effective conductivity is omitted here due to the negligible propagation loss. To further illustrate the superiority of the concept of equivalent graphene, next, we construct the anisotropic metamaterials by stacking EGPMs and the dielectric with relative permittivity $\varepsilon_r$ periodically, as illustrated by **Figure 2**. The distance between the adjacent EGPMs is $d$. Using the Floquet Theory, we derive the dispersion relation of the constructed anisotropic metamaterials and obtain all the components of the effective permittivity tensor. For a monolayer EGPM, its transfer matrix has the form

$$T_1 = \begin{bmatrix} 1 & 0 \\ \sigma_m & 1 \end{bmatrix}. \tag{7}$$

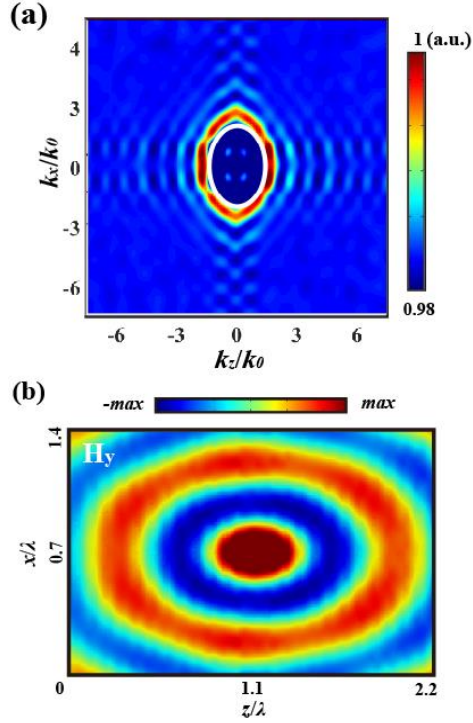

**Figure 4**. (a) Theoretical and practical EFCs of the constructed EMM with $\varepsilon_e = 1.8$ and $\varepsilon_r = 4.3 - 0.01i$. (b) The magnetic field distribution in the EMM excited by a magnetic dipole.

Similarly, the transfer matrix of the dielectric with thickness $d$ has the form

$$T_d = \begin{bmatrix} \cos(k_\perp d) & jZ_0 \sin(k_\perp d) \\ \dfrac{j}{Z_0} \sin(k_\perp d) & \cos(k_\perp d) \end{bmatrix}, \tag{8}$$

where $Z_0$ represents the impedance in free space and $k_\perp = \sqrt{\varepsilon_r k_0^2 - k_x^2}$. Then, the transfer matrix of the repeated unit composed of an EGPM sheet and a dielectric layer can be written as

$$T_{unit} = T_\sigma T_d = \begin{bmatrix} \cos(k_\perp d) & jZ_0 \sin(k_\perp d) \\ \dfrac{j}{Z_0} \sin(k_\perp d) + \sigma_m \cos(k_\perp d) & j\sigma_m Z_0 \sin(k_\perp d) + \cos(k_\perp d) \end{bmatrix}. \tag{9}$$

The tangential electromagnetic wave at $z+d$ obtained by Floquet Theory and Transfer Matrix Method can be respectively written as

$$\begin{bmatrix} E_t(z+d) \\ H_t(z+d) \end{bmatrix} = e^{-jk_z d} \begin{bmatrix} E_t(z) \\ H_t(z) \end{bmatrix}, \tag{10a}$$

$$\begin{bmatrix} E_t(z+d) \\ H_t(z+d) \end{bmatrix} = T_{unit} \begin{bmatrix} E_t(z) \\ H_t(z) \end{bmatrix}, \tag{10b}$$

where $k_z$ represents the Floquet wave vector. Therefore, the wave vector dispersion relation of multilayer anisotropic metamaterials can be obtained from the solution of the equation

$$\det\left|T_{unit} - e^{-jk_z d} I\right| = 0, \tag{11}$$

where $\mathbf{I}$ represents the identity matrix. Substituting Equation (9) into Equation (11), we can obtain that

$$\cos(k_z d) = \cos(k_\perp d) + j \frac{\sigma_m k_\perp}{2\omega \varepsilon_0 \varepsilon_r} \sin(k_\perp d). \tag{12}$$

In this work, the distance $d$ between the adjacent EGPM sheets is deep subwavelength, so the inequation $k_z d \ll 1$ and $k_\perp d \ll 1$ always hold. Under these conditions, we substitute the equivalent infinitesimal $\cos(k_z d) = 1 - \frac{1}{2}(k_z d)^2$, $\cos(k_\perp d) = 1 - \frac{1}{2}(k_\perp d)^2$, and $\sin(k_\perp d) = k_\perp d$ into Equation (12) and the dispersion relation finally arrives at

$$(k_z d)^2 = (k_\perp d)^2 - j\frac{\sigma_m k_\perp^2 d}{\omega \varepsilon_0 \varepsilon_r}. \qquad (13)$$

Defining $\varepsilon_e = \varepsilon_r - j\frac{\sigma_m}{\omega \varepsilon_0 d}$, the dispersion relation is finally simplified as

$$\frac{k_z^2}{\varepsilon_e} + \frac{k_{//}^2}{\varepsilon_r} = k_0^2. \qquad (14)$$

The whole system can be treated as a homogeneous medium with anisotropic parameters $\varepsilon_e$ and $\varepsilon_r$. When we adjust the effective conductivity $\sigma_m$ and the distance $d$ so that $\varepsilon_e < 0$ or $\varepsilon_e > 0$ meets, hyperbolic or elliptical EFCs can be efficiently obtained, respectively. To investigate the achievable range of $\varepsilon_e$, we calculated the effective permittivity $\varepsilon_e$ versus $\sigma_m$ and $d$ under $\varepsilon_r = 3.5 - 0.01i$ at 2GHz. As can be seen in **Figure 3(c)**, a large effective permittivity coverage from −221 to 2.5 can be obtained when we adjust the effective conductivity $\sigma_m$ from $-25i$ mS to $-1i$ mS and the distance $d$ from 2 mm to 18 mm. The lower limit of the coverage can be smaller by narrowing the distance $d$ or decreasing the effective conductivity $\sigma_m$.

## 4. Design and Fabrication of Anisotropic Metamaterials

### 4.1. Elliptical Metamaterial

To demonstrate the practical effect of the proposed method, firstly, we design an elliptical metamaterial(EMM) with positive $\varepsilon_e$ and $\varepsilon_r$. To achieve this goal, we set the structure parameters with $p_y = 17.5$ mm, $p_x = 10$ mm, $d = 10$ mm, $w = 2.5$ mm, $b = 16.5$ mm, and $\varepsilon_r = 4.3 - 0.01i$ so the effective permittivity in x direction can be obtained with $\varepsilon_e = 1.8$. Note that, all of the results calculated and shown below are obtained at $f = 2$ GHz. The white ellipse in **Figure 4(a)** displays the theoretical EFC of this EMM with $k_z^2/\varepsilon_e + k_x^2/\varepsilon_r = 1$. After the theoretical analysis, we construct the practical EMM and numerically validate its functions. In this model, the electromagnetic waves are excited by a magnetic dipole at the center of the EMM. As shown by the magnetic field distribution in **Figure 4(b)**, with the model surrounded by perfect absorption boundary conditions, an elliptical EMW propagation is visually obtained. The long axis of the isophase figure in Figure 4(b) is vertical to that of theoretical EFC in Figure 4(a) because the wavelength and the wave vector are inversely proportional. To quantitatively evaluate the accuracy of the constructed EMM, we calculate its EFC by applying a 2D spatial Fourier transform on the magnetic field excited by the magnetic dipole. The result is plotted in Figure 4(a) by pseudo-color image. The consistency between the theoretical and practical EFCs in Figure 4(a) can effectively validate the effectiveness of the method. The inconspicuous ripples in Figure 4(a) are attributed to the imperfections of the absorption boundary condition and the truncation effect of the Fourier transform. It's worth noting that the legend of the pseudo-color image of the practical EFC starts at 0.98 and ends at 1 (a.u.). The huge background noise of the wave vector below 0.98 arises from the high magnetic field intensity of the magnetic dipole whose wave vector uniformly covers the entire wave vector space. This noise can be eliminated by only calculating the EFC of the scattering magnetic field.

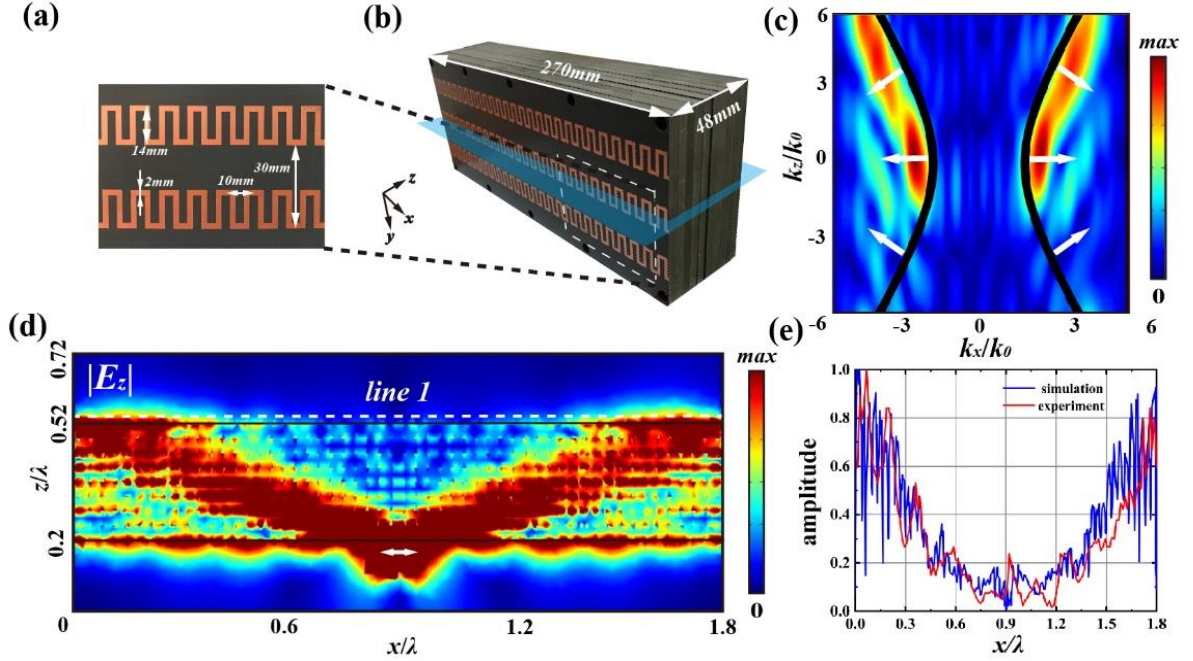

**Figure 5**. (a) The exact structure parameters of the HMM, where $p_y = 30$ mm, $p_x = 10$ mm, $d = 6$ mm, $w = 2$ mm, and $b = 14$ mm. (b) Picture of the fabricated HMM. (c) Theoretical and practical EFCs of the HMM with $\varepsilon_e = -10$ and $\varepsilon_r = 3.5 - 0.01i$. The white arrows vertical to the hyperbola represent the directions of group velocity. (d) The electric field $\mathbf{E}_z$ distribution in the HMM excited by an electric dipole, corresponding to the blue plane in (b). Line 1 represents the path the monopole antenna goes along to measure the $\mathbf{E}_z$ distribution. (e) The numerical and experimental $\mathbf{E}_z$ distributions corresponding to line 1 in (d).

### 4.2. Hyperbolic Metamaterial

When the effective permittivity $\varepsilon_e$ goes from positive to negative, a topological transition of EFCs from ellipse to hyperbola will occur. To further illustrate the superiority of the method, next, we design and fabricate an anisotropic HMM with negative $\varepsilon_e$. By confirming the structure parameters with $p_y$ = 30 mm, $p_x$ = 10 mm, $d$ = 6 mm, $w$ = 2 mm, $b$ = 14 mm, and $\varepsilon_r$ = 3.5 − 0.01$i$, a theoretical hyperbolic EFC with $\varepsilon_e$ = − 10 can be obtained, as plotted by the black curve in **Figure 5(c)**. The white arrows vertical to the hyperbola represent the directions of group velocity. **Figure 5(b)** shows the picture of the fabricated HMM prototype, with its detailed structure parameters marked in **Figure 5(a)**. To validate its functions, we calculate the electric field distribution under the excitation of a parallel electric dipole 5 mm in front of the HMM. As shown by the $\mathbf{E}_z$ distribution in **Figure 5(d)** corresponding to the blue plane in Figure 5(b), the electromagnetic waves radiated from the electric dipole propagate only towards two specific symmetrical directions. This is because the group velocity direction is perpendicular to the dispersive curve so the spatial harmonics with high $k_z$ only have two preferred directions, as indicated by the white arrows in Figure 5(c). The directional propagation can effectively validate the functions of the designed HMM. To further quantitatively evaluate the accuracy of the constructed EMM compared with theory, Similarly, a 2D spatial Fourier transform is also applied on the electric field distribution of the HMM in Figure 5(d) to obtain the practical EFC. The result is illustrated in Figure 5(c) by the rainbow-color image, where the practical EFC is almost the same as the top half of the theoretical one. The deviation between them arises from the reflection at the air-HMM interface and the non-ideal absorption boundary condition around the HMM. Compared with the theoretical curve, the bottom half is missing because only the electromagnetic waves radiated by the dipole with positive $k_z$ are coupled into the HMM. The complete EFC can be obtained by placing another same HMM on the other side of the dipole. For the experimental investigation, a two-port AV3656A Vector Network Analyzer(VNA) is used with its ports connected with two monopole antennas. One antenna is used to excite the prototype, while the other one is used to measure the response by a scanning process. Similar to the configuration of the simulation scenario above, the emitting antenna is placed in front of the center of the HMM with its probe parallel to the facade. The absorbing materials are placed around the HMM to absorb the scattering waves and imitate infinite space. The white dotted line in Figure 5(d) displays the route the receiving antenna goes along to measure the $\mathbf{E}_z$ distribution. The scanning step of the receiving monopole antenna, which is controlled by a 2D electric-controlled stage, is 2 mm, as illustrated by **Figure 8(f)**. The distance between the line and the HMMs is 3 mm. As can be seen from **Figure 5(e)**, the red line which is measured by the monopole antenna matches well the simulated blue one, validating the functions of the constructed HMM.

In the designed HMMs, selective directional propagation, which is also called photonic spin Hall effect, can be realized by adjusting the handedness of the circularly polarized electric dipoles.[39] The discovery of PSHE boosts the field of spintronics and it may find applications in quantum information technologies, imaging, and biosensing.[39-42] When we place a linearly polarized electric dipole in front of the HMMs, directional propagation will be observed with trumpet-shaped field distribution [39]. However, as illustrated by **Figure 6(a)**, if we replace the linearly polarized dipole with the circularly polarized one whose dipole moment is $\hat{\boldsymbol{p}} = \hat{\boldsymbol{x}} - i\hat{\boldsymbol{z}}$, the selective directional propagation that only one side of the original trumpet-shaped space is excited will occur. Similarly, in **Figure 6(b)**, the other side will be excited by changing the handedness of the electric dipole. This phenomenon is attributed to the break of symmetry caused by the circularly polarized dipoles. For the linearly polarized dipole with dipole moment $\hat{p} = \hat{x}$, the excited electric field in the HMM will propagate towards the two specified directions which are symmetrical with $yz$ plane, as illustrated by Figure 5(d). Besides, the amplitude and phase of the magnetic field in both directions are also symmetrical. For the linearly polarized dipole with dipole moment $\hat{p} = \hat{z}$, the difference is that the amplitude of the magnetic field in the both directions are symmetric while the phase is antisymmetric. Therefore, for the circularly polarized dipole which can be seen as the linear superposition of the two orthogonal dipoles, PSHE will be observed by the constructive interference in one direction while destructive interference in the other direction. To examine the performance of the selective directional propagation, we perform a microwave experiment on the fabricated HMM. Similar to the measuring configuration above, an emitting monopole antenna is used to excite the prototype while the other one is used to measure the response. To imitate the circularly polarized dipole, two measurements are performed with the emitting antenna vertical and parallel to the facade, respectively. **Figure 6(d)** and **6(e)** illustrate the numerical electric field $\mathbf{E}_z$ distributions excited by the electric dipoles with dipole moment $\hat{\boldsymbol{p}} = \hat{\boldsymbol{x}} - i\hat{\boldsymbol{z}}$ and $\hat{\boldsymbol{p}} = \hat{\boldsymbol{x}} + i\hat{\boldsymbol{z}}$, respectively. The white circles with arrows 5 mm in front of the HMMs represent the emitting monopole antennas. As illustrated by Figure **6(c)**, we plot the numerical and experimental $\mathbf{E}_z$ distribution corresponding to line 1 in Figure 6(d) and line 2 in Figure 6(e), where evident selective directional propagation can be observed. The consistency between numerical and experimental results indicates the good performance of the HMMs on PHSE. The slight deviation between the simulated and experimental results is attributed to the tolerances of fabrication, arrangement and measurement, and the limited absorbing effect of the absorbing materials. In Figure 6(d) and 6(e), evanescent waves are excited at the air-HMM interfaces because an ideal and local boundary that can support the Eigen surface wave is effectively formed considering the positive and negative effective permittivity on both sides.

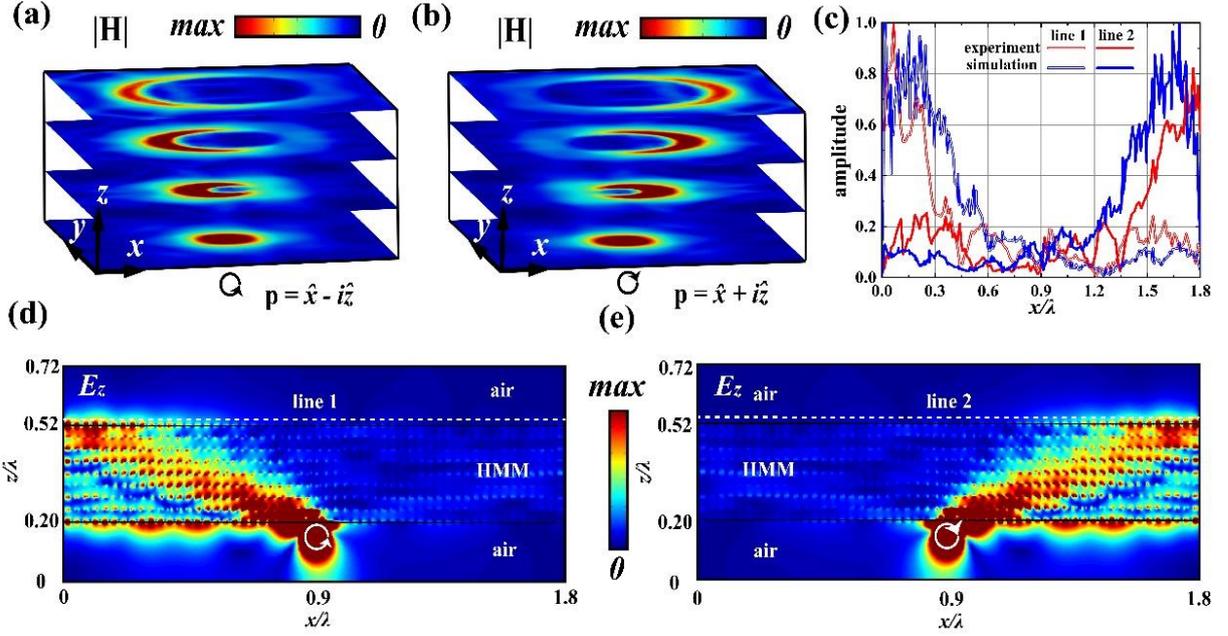

**Figure 6**. (a) and (b) are schematics of PSHE excited by electric dipoles with respective dipole moment $\hat{p} = \hat{x} - i\hat{z}$ and $\hat{p} = \hat{x} + i\hat{z}$. (c) The numerical and experimental $E_z$ distributions correspond to line 1 in (d) and line 2 in (e). (d) and (e) are the respective 2D $E_z$ distributions correspond to the situations in (a) and (b). lines 1 and 2 are the paths the monopole antenna goes along to measure the $E_z$ distribution. The distance between the lines and the HMMs is 3 mm and the scanning step of the monopole antenna is 2 mm.

## 5. All-metal Hyperbolic Metamaterial

The formula of the effective conductivity $\varepsilon_e = \varepsilon_r - j\dfrac{\sigma_m}{\omega\varepsilon_0 d}$ tell us that arbitrary $\varepsilon_e$ smaller than $\varepsilon_r$ can be theoretically realized by designing the anisotropic metamaterials with appropriate $\sigma_m$ and $d$. It means that the permittivity of the dielectric is not the only determining factor of $\varepsilon_e$ and arbitrary natural materials, even including air, can be selected to construct the HMMs according to the proposed method. To demonstrate this advantage, next, we design a all-metal HMM with $\varepsilon_e = -1$ and $\varepsilon_r = 1$, and verify its functions through negative refraction. The corresponding structure parameters are $p_y = 17.5$ mm, $p_x = 10$ mm, $d = 20$ mm, $w = 2$ mm, and $b = 15$ mm. Different from the previous situation, the EGPMs cannot be directly stacked since there is no solid material to support them. Here, we use the thin metal strips with a width of 3 mm to connect the snakelike EGPMs, as illustrated by **Figure 7(a)**. The influence of these upright metal strips can be ignored because the incident waves are TM polarization whose electric field components are completely vertical to the metal strips. In this case, no electric current is excited at the surface of the metal strips. Four horizontal metal posts are then used to concatenate each piece of EGPM together. Compared with the conventional dielectric-based HMM, the all-metal structure of the HMM make it qualified to work in high-power scenarios. Besides, the overall loss can be further decreased for the absence of lossy dielectric. In **Figure 7(b)**, we plot the theoretical EFCs of air and the all-metal HMM. The red arrow represents the wave vector of the incident Gaussian wave with $\theta_i = 30°$, and the green arrow, which is vertical to the hyperbola, represents the theoretical direction of group velocity in the all-metal HMM according to tangential wave vector conservation. In this case, negative refraction is achieved in the HMM with a refraction angle $\theta_r = -24°$. **Figure 7(c)** illustrates the electric field $\mathbf{E}_z$ distribution of this situation, where the boundaries of the all-metal HMM are represented by the white dotted lines and the length ($z$ direction) and width ($x$ direction) of the all-metal HMM is $3.2\lambda$ and $1.3\lambda$, respectively. The waist width of the incident Gaussian wave is $2.9\lambda$. The corresponding red and green arrows in Figure 7(c) are parallel to those in Figure 7(b), indicating an ideal performance of the all-metal HMM compared with theory. Besides, the transmissivity, which is defined as the power ratio between the transmissive and incident waves, can reach up to 83.4%. The high efficiency is mainly attributed to the same medium between the HMM and the surrounding.

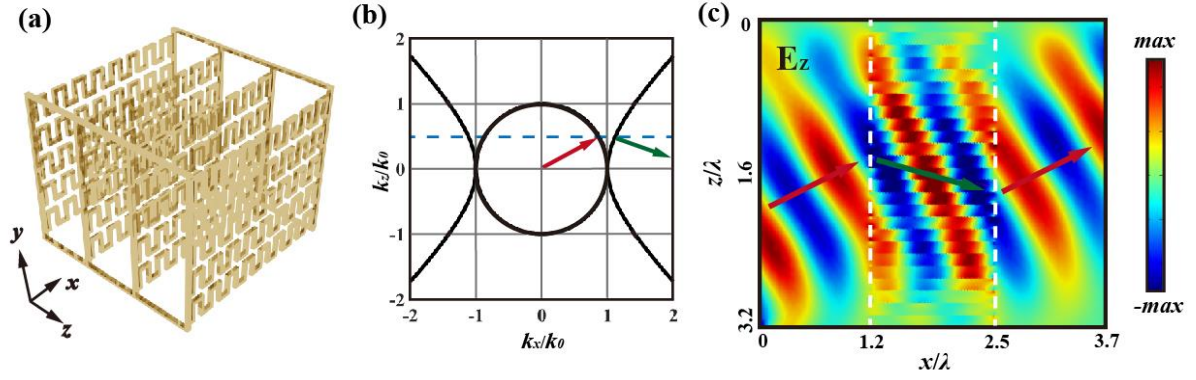

**Figure 7.** (a) Schematic of the all-metal HMMs. (b) EFCs of air and the all-metal HMMs with $\varepsilon_e = -1$. The red arrow represents the wave vector of the incident wave with $\theta_i = 30°$ and the green arrow represents the theoretical refracted wave vector in the all-metal HMM according to tangential wave vector conservation. (c) Electric field $\mathbf{E}_z$ distribution of anomalous reflection when a Gaussian wave passes through the all-metal HMM. The white dotted lines represent the boundaries between air and HMM.

## 6. Hyperlens with Extremely Small Effective Permittivity

HMM is a good candidate to realize subwavelength resolution imaging because it allows the propagation of evanescent waves with large lateral wave vector components. However, to construct a practical hyperlens, the permittivity in transmission direction should be as small as possible so that the EMWs with different lateral wave vector components can have the same group velocity direction. To realize this, the width $b$ and the distance $d$ are reduced to 6 mm and 2 mm, respectively, and the resulting permittivity in transmission direction can reach down to $\varepsilon_e = -510$. The effective permittivity $\varepsilon_e$ can be further reduced by decreasing $b$ and $d$. However, this effort has a limited effect on improving performance. The other parameters of the HMM are set with $p_y = 12$ mm, $p_x = 13.4$ mm, $w = 4$ mm, and $\varepsilon_r = 3.5 - 0.01i$, as shown by the inset in **Figure 8(c)**. We plot the theoretical dispersion curve of the hyperlens in **Figure 8(a)**, where the white arrows representing the group velocity directions are almost parallel to each other. To obtain the practical EFC of the hyperlens, we place a magnetic dipole at the center and calculate its magnetic field distribution, as illustrated in Figure 8(c). The practical EFC of the hyperlens is calculated based on the magnetic field distribution and plotted in Figure 8(a). The ideal overlap of the theoretical and practical EFCs verifies the validity of the hyperlens. To explore the limit of its imaging resolution, we fabricate the practical hyperlens and perform a microwave experiment on it, as illustrated by **Figure 8(e)**. Figure 8(f) illustrates the main measurement configuration. One of the monopole antenna connected to the port 1 of the VNA is fixed 2 mm behind the hyperlens, while the other one connected to the port 2 and fixed in the electric-controlled stage is used to measure the $\mathbf{E}_z$ distribution 2 mm in front of the hyperlens. The scanning step of the receiving monopole antenna is 1 mm. **Figure 8(b)** illustrates the numerical electric field distribution of the entire imaging process corresponding to the experiment configuration. The white dotted line represents the path the receiving antenna goes along to measure the image of the dipole source. It's clear to see that the EMW propagates in a self-collimating manner and no evident diffraction occurs so the sources can be ideally transferred to the image side with little distortion. The numerical and experimental electric field distribution of the target in the imaging side are plotted in **Figure 8(d)**. An ultimate deep subwavelength imaging resolution of 0.05λ (λ is the operating wavelength), which is estimated by full-width at half-maximum (FWHM) of one peak in the 1D intensity of electric field distribution, is experimentally obtained. The inset illustrates the 2D image of the electric dipole source obtained in the imaging side. Owing to the advantage of low loss, the value is much better than that of the previous hyperlens.[12-14,43-45] The measured results demonstrate the effectiveness of the proposed method under extreme parameter conditions. Compared with the simulated resolution of 0.035λ, the experimental result can be further promoted by reducing the tolerances of fabrication and improving the accuracy of the measurement.

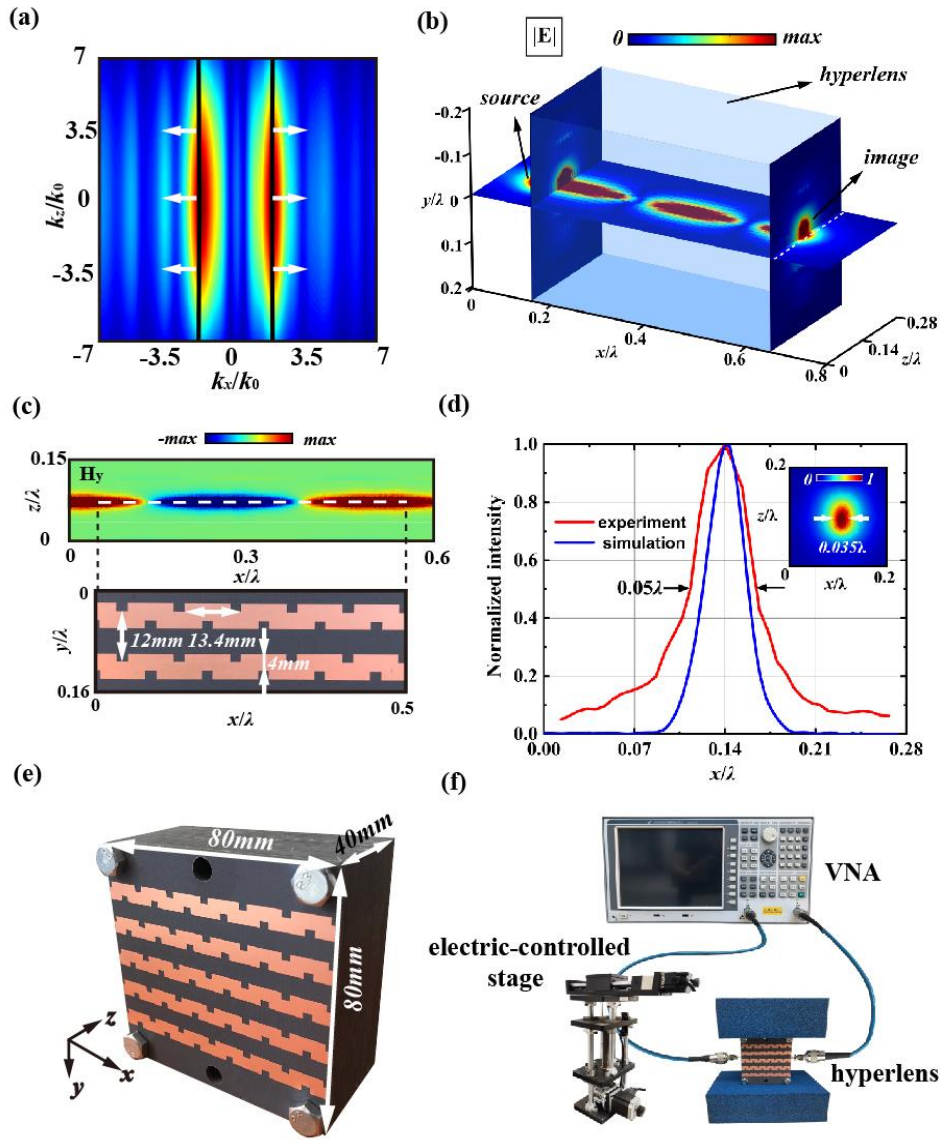

**Figure 8.** (a) Theoretical and practical EFCs of the hyperlens with $\varepsilon_e = -510$. The white arrows represent the directions of the group velocity. (b) The electric field distribution of the entire imaging process. (c) Numerical magnetic field distribution of the hyperlens excited by a magnetic dipole. The inset shows the exact structure parameters of the hyperlens. (d) The numerical and experimental electric field $\mathbf{E}_z$ distribution of targets in the imaging side. (e) Picture of the fabricated hyperlens. (f) The schematic of the main measurement configuration.

## 7. Conclusion

In this work, we propose the concept of equivalent graphene and demonstrate that it can be modeled as a sheet with effective conductivity $\sigma_m$. Based on this concept, elliptical and hyperbolic dispersion relation can be theoretically obtained by stacking the equivalent graphene metasurfaces periodically. To illustrate the superiority of the proposed concept and method, we design the layered anisotropic metamaterials and demonstrate their electromagnetic properties including elliptical propagation, directional propagation, PSHE, negative refraction, and non-diffraction propagation. Our study may open up new routes in designing plasmonics metamaterials such as focusing lens, dispersion-dependent directional couplers, and other practical microwave and terahertz devices.

## 8. Experimental Section

*Fabrication of the samples*

The samples are constructed by stacking the EGPMs periodically. For the EGPMs of the HMM and the hyperlens, the sizes are 270 mm × 110 mm × 6 mm and 80 mm × 80 mm × 2mm, respectively. A certain number of holes with a diameter of 7 mm are punched on the edge of the EGPMs for the convenience of fixing them together. The EGPMs are made of F4B and copper. The copper with a thickness of 0.017 mm is coated on the surface of the F4B by the chemical deposition. Antioxidant technology is also applied to the surface of the EGPMs to protect the samples.

*The experimental setup*

To obtain the accurate results of the electric field distribution, the monopole antenna connected with one port of the VNA is fixed in front of the samples by the holder. The length of the probe of the monopole antenna is 3mm, short enough to be equivalent to an electric dipole. Another monopole antenna is fixed on the 2D electric-controlled stage whose position and scanning step can be controlled by the controller and the computer. To imitate infinite space, enough absorbing materials are placed around the samples and the monopole antennas. After connecting the VNA and the controller with the computer, the electric field data at the desired position can be obtained by entering the corresponding instructions on the computer.

**Acknowledgements**

This work is sponsored by National Key Research and Development Program (2019YFA0210203) and National Natural Science Foundation of China (NSFC) (61971013).

Received: ((will be filled in by the editorial staff))
Revised: ((will be filled in by the editorial staff))
Published online: ((will be filled in by the editorial staff))